\documentclass[10pt,aps,prl,twocolumn,superscriptaddress,showpacs,preprintnumbers]{revtex4-1}

\usepackage{amsmath,amssymb,amsfonts}
\usepackage{graphicx,color}
\usepackage{verbatim}
\usepackage{float}
\usepackage{dcolumn}
\usepackage{natbib}
\usepackage{bm}
\usepackage{hyperref}
\usepackage{soul}

\newcommand{\ie}{\emph{i.e.}}
\newcommand{\eg}{\emph{e.g.}}

\newcommand{\add}[1]{\textcolor{black}{#1}}

\begin{document}

\title{Enhanced capital-asset pricing model for the reconstruction of bipartite financial networks}

\author{Tiziano Squartini}
\affiliation{IMT School for Advanced Studies, Piazza S.Francesco 19, 55100 Lucca - Italy}
\author{Assaf Almog}
\affiliation{Instituut-Lorentz for Theoretical Physics, Leiden Institute of Physics, University of Leiden, Niels Bohrweg 2, 2333 CA Leiden - The Netherlands}
\author{Guido Caldarelli}
\affiliation{IMT School for Advanced Studies, Piazza S.Francesco 19, 55100 Lucca - Italy}
\affiliation{Istituto dei Sistemi Complessi (ISC)-CNR UoS Universit\`a ``Sapienza'', Piazzale Aldo Moro 5, 00185 Rome - Italy}
\author{Iman van Lelyveld}
\affiliation{De Nederlandsche Bank, PO box 98, 1000 AB Amsterdam - The Netherlands}
\affiliation{VU University, De Boelelaan 1105, 1081 HV Amsterdam - The Netherlands}
\author{Diego Garlaschelli}
\affiliation{Instituut-Lorentz for Theoretical Physics, Leiden Institute of Physics, University of Leiden, Niels Bohrweg 2, 2333 CA Leiden - The Netherlands}
\author{Giulio Cimini}
\affiliation{IMT School for Advanced Studies, Piazza S.Francesco 19, 55100 Lucca - Italy}
\affiliation{Istituto dei Sistemi Complessi (ISC)-CNR UoS Universit\`a ``Sapienza'', Piazzale Aldo Moro 5, 00185 Rome - Italy}

\begin{abstract}
Reconstructing patterns of interconnections from partial information is one of the most important issues in the statistical physics of complex networks. 
A paramount example is provided by financial networks. In fact, the spreading and amplification of financial distress in capital markets is strongly affected by 
the interconnections among financial institutions. Yet, while the aggregate balance sheets of institutions are publicly disclosed, 
information on single positions is mostly confidential and, as such, unavailable. 
Standard approaches to reconstruct the network of financial interconnection produce unrealistically dense topologies, leading to a biased estimation of systemic risk. 
Moreover, reconstruction techniques are generally designed for monopartite networks of bilateral exposures between financial institutions, 
thus failing in reproducing bipartite networks of security holdings (\eg, investment portfolios). 
Here we propose a reconstruction method based on constrained entropy maximization, tailored for bipartite financial networks. 
Such a procedure enhances the traditional {\em capital-asset pricing model} (CAPM) and allows to reproduce the correct topology of the network. 
We test this ECAPM method on a dataset, collected by the European Central Bank, of detailed security holdings of European institutional sectors over a period of six years (2009-2015). 
Our approach outperforms the traditional CAPM and the recently proposed MECAPM both in reproducing the network topology and in estimating systemic risk 
\add{due to fire-sales spillovers. In general, ECAPM can be applied to the whole class of weighted bipartite networks described by the fitness model.}

\end{abstract}

\maketitle 

\section{Introduction}\label{intro}

The recent financial crises have highlighted the importance of correctly evaluating systemic risk in financial markets, 
by explicitly considering the role played by financial interconnections 
(both as direct exposures through bilateral contracts and indirect exposures through common assets holding~\cite{Eisenberg2001,Bonanno2003,Gai2010,Haldane2011,Acemoglu2015,Cimini2016}) 
in the spreading of financial distress~\cite{Allen2000,Lau2009,Brunnermeier2009a,Krause2012,Battiston2015price}. 
Thus, characterising the underlying network structure of financial systems has become not only a scientific 
but also an institutional priority~\cite{Iori2006,Elsinger2006,Nier2008,Bardoscia2016X,Battiston2016}. 
However, while financial institutions have to disclose their aggregated exposures (eventually split into major categories), 
detailed data on their single positions are generally confidential and, thus, unaccessible. 
For this reason, reconstructing the network structure from the available data 
has recently attracted much attention~\cite{Wells2004,Upper2011,Mistrulli2011,Mastromatteo2012,Baral2012,Drehmann2013,Halaj2013,Anand2014,Montagna2014,Peltonen2015,Cimini2015a,Cimini2015b,Gandy2015X}. 
Yet, while several efforts have been devoted so far to reconstruct {\em monopartite} networks of bilateral exposures between financial institutions, 
methods to reconstruct the {\em bipartite} structure of assets ownership by such institutions are still little explored~\cite{DiGiangi2015}. 
Assessing the portfolio composition of both financial actors and the institutional sectors they belong to is, however, crucial in terms of systemic stability. 
In fact portfolio overlaps have the potential to trigger {\em fire sales}, \ie, downward spirals for illiquid asset prices 
due to self-reinforcing sell orders~\cite{Shleifer2010,Caccioli2014,Greenwood2015,Cont2016,Gualdi2016X}. 
Fire sales spillovers are also dangerous because they create incentives to hoard liquidity, thus activating a spiral 
potentially leading to a complete freeze of the financial system~\cite{Adrian2010,Diamond2010,Berrospide2012,Gale2013,Acharya2013,Gabrieli2014}. 

Reconstruction of monopartite networks of direct exposures between financial institutions (such as interbank markets) 
is typically pursued through dense maximum-entropy techniques that use the available information as constraints to be satisfied~\cite{Wells2004,Upper2011}. 
The major drawback of these approaches is however that of generating unrealistically (almost) fully-connected topologies which, in turn, 
\add{were shown to underestimate systemic losses due to counterparty risk}~\cite{Mistrulli2011,Mastromatteo2012}. 
Several techniques have thus been developed to obtain \emph{sparse} reconstructions~\cite{Mastromatteo2012,Drehmann2013,Halaj2013,Anand2014} by inferring connection probabilities 
via the aggregated exposures~\cite{Montagna2014}.
While these approaches either explore the whole range of possible network densities or directly make use of {\em ad-hoc} density values, 
a recently proposed {\em bootstrapping} algorithm allows one to obtain an accurate guess of the true network density relying on small, informative, subsets of nodes~\cite{Musmeci2013,Cimini2015-0}. 
The latter approach has then been developed into a statistically-grounded reconstruction procedure, which assesses connection probabilities 
through a {\em configuration model} (CM) \cite{Park2004,Squartini2011} tailored on the estimated density and induced by the aggregated exposures used as {\em fitnesses} \cite{Caldarelli2002,Garlaschelli2004}. 
In order to evaluate the weights of connections, the approach can be either complemented with a {\em degree-corrected gravity model} (dcGM)~\cite{Cimini2015a} 
or with an {\em enhanced} CM (ECM~\cite{Mastrandrea2014})~\cite{Cimini2015b}. 
As pointed out by recent studies \cite{Mazzarisi2017,Anand2017}, this method outperforms other existing (probabilistic) reconstruction techniques. 

Differently from interbank markets, networks of institutional portfolio holdings are systems composed by two classes of nodes: institutions (or sectors) and financial instruments. 
In this case a reconstruction method has to account for the bipartite structure of the system, since connections are allowed only between nodes belonging to different classes. 
Standard maximum entropy techniques have been extended also to deal with such systems, generating candidate configurations according to the {\em capital asset pricing model} (CAPM)---for which 
investors choose their portfolio so that each position on a given asset is proportional to that asset's market capitalization~\cite{Lintner1965,Mossin1966}. 
Recently, a cross-entropy maximization technique constraining weight expectations to match CAPM values (MECAPM) has also been formalized~\cite{DiGiangi2015}. 
Yet, both approaches predict very dense configurations, thus failing in reproducing the real network structure, \add{and eventually underestimating systemic risk 
arising from fire sales spillovers}~\cite{Caccioli2014} (although this may depend on the metric chosen~\cite{DiGiangi2015}).

In this paper we aim at overcoming the density issue for bipartite networks reconstruction, by defining an {\em enhanced} CAPM (ECAPM), which extends the {\em fitness-induced} CM 
introduced for monopartite networks~\cite{Cimini2015a} to the bipartite case. In a nutshell, the method consists of a two-step inference procedure:
first, links presence is assessed through a {\em bipartite} CM (BiCM)~\cite{Saracco2015} calibrated using the aggregated balance sheet data as fitnesses; 
then, weights are estimated via the {\em degree-corrected} CAPM. 
To validate our method, we use a unique dataset of security holdings by institutional sectors in Europe (in particular, we have the detailed exposures of long-term security bonds between sectors). 
Since we have full information on these data (which is confidential and not publicly available), we are able to precisely assess the accuracy of our reconstruction method.

\section{Method}

\add{\bf Notation} --- We start by introducing the notation that is used throughout the paper. 
Assets ownership data are represented as a weighted, undirected, bipartite network $G_0$ of $N$ investors portfolios and $M$ assets. 
The generic element of the $N\times M$ biadjacency matrix, $w_{i\alpha}$, is the value of asset $\alpha$ held by portfolio $i$. 
The aggregated balance sheet information is thus given by the nodes {\em strengths}, 
namely the market value of portfolio $i$, $V_i=\sum_\alpha w_{i\alpha}$, and the market capitalization of asset $\alpha$, $C_\alpha=\sum_i w_{i\alpha}$. 
Connection patters are instead described by the $N\times M$ binary biadjacency matrix, whose generic element $a_{i\alpha}$ equals 1 if $w_{i\alpha}>0$ and 0 otherwise. 
Nodes {\em degrees} are then given by the number of different assets held by each portfolio $i$, $k_i=\sum_\alpha a_{i\alpha}$, and by the number of investors in each asset $\alpha$, 
$d_\alpha=\sum_i a_{i\alpha}$. $W=\sum_{i\alpha}w_{i\alpha}$ denotes the total economic value of the system, 
and $L=\sum_{i\alpha}a_{i\alpha}$ is the total number of connections (ownership relations) in the network. 

\medskip

\add{{\bf Problem statement} --- Suppose to have limited information on $G_0$: we know the whole strength sequences $\{V_i\}_{i=1}^N$ and $\{C_\alpha\}_{\alpha=1}^M$ and the total number of links $L$ 
(note that strengths are usually publicly available and, thus, easy to access, whereas $L$ is more difficult to obtain and is typically proxied through density sampling 
or bootstrapping techniques~\cite{Musmeci2013,Squartini2017}). In this situation, our goal is to find the optimal estimate for $X(G_0)$, the value of a generic property $X$ measured on the real network $G_0$ 
on the basis of the information available.}

\medskip

\add{The ECAPM network reconstruction procedure works in two main steps: it supposes $G_0$ as drawn from an appropriate ensemble $\Omega$ 
of bipartite graphs, defined by a {\em fitness-induced} BiCM for links presence and by a {\em degree-corrected} CAPM for links weight. 
In what follows, we will label with the tilde symbol $\sim$ quantities of networks drawn from $\Omega$.}

\medskip

{\bf Topology reconstruction} --- If we knew the degree of each node in the network, we could use the BiCM \cite{Saracco2015} to directly generate $\Omega$ 
as the ensemble of binary, undirected bipartite networks satisfying the degree constraints 
$\langle k_i \rangle_{\Omega}=k_i(G_0)$, $i=1\dots N$ and $\langle d_\alpha\rangle_{\Omega}=d_\alpha(G_0)$, $\alpha=1\dots M$. 
This would imply a probability distribution over $\Omega$ defined by a set of Lagrange multipliers $\{x_i\}_{i=1}^N$ and $\{y_\alpha\}_{\alpha=1}^M$ 
(one for each node) associated to the degrees, so that the connection probability between nodes $i$ and $\alpha$ is
\begin{equation}\label{eq:prob}
p_{i\alpha}\equiv \langle\tilde a_{i\alpha}\rangle_\Omega=\frac{x_iy_\alpha}{1+x_iy_\alpha},
\end{equation}
independently for all nodes pairs. Here, however, we are studying the case where node degrees are unknown, yet we know the total number $L$ of links. 
Thus, while we cannot directly use the BiCM, we can resort to the {\em fitness} ansatz, which assumes the network topology to be determined by intrinsic node properties named fitnesses. 
This approach has been successfully used in the past to model several economic and financial networks, by assuming a proportionality 
between fitnesses and Lagrange multipliers~\cite{Garlaschelli2004,Garlaschelli2005,DeMasi2006}. 
Thus, as in the case of monopartite financial networks~\cite{Cimini2015a,Cimini2015b}, we assume the strengths (which we know) to represent node-specific fitnesses 
and be linearly proportional to the BiCM Lagrange multipliers (induced by degrees): $x_i\equiv \sqrt{z_V} V_i$, $\forall i$ and $y_\alpha\equiv \sqrt{z_C} C_\alpha$, $\forall\alpha$. 
Thanks to this assumption, our task reduces to determining only one proportionality constant, which is obtained by equating the ensemble average of the total number of links 
with the (known) total number $L$ of links of $G_0$:
\begin{equation}\label{eq:L} 
\langle L \rangle_{\Omega} \equiv \sum_i \sum_\alpha \frac{zV_iC_\alpha}{1+zV_iC_\alpha} = L(G_0),
\end{equation} 
where we have defined $z=\sqrt{z_Vz_C}$. Eq.~(\ref{eq:L}) is an algebraic equation in $z$ with a single positive solution, which is then used to estimate the linking probabilities of eq.~(\ref{eq:prob}):
\begin{equation}
 p_{i\alpha}\equiv\langle \tilde a_{i\alpha}\rangle_{\Omega} = \frac{zV_iC_\alpha}{1+zV_iC_\alpha}\qquad\forall(i,\alpha).
\label{eq:prob_plus}
\end{equation}
By preserving the network density, the topology predicted by eq.~(\ref{eq:prob_plus}) generally differs substantially 
from the one characterizing the fully, or very densely, connected configurations predicted by the CAPM and MECAPM, respectively. 

\medskip

{\bf Weights reconstruction} --- If the information on nodes degrees were accessible, the formal approach to obtain a weighted reconstructed network would prescribe to use the ECM to maximize entropy, 
constraining both degrees and strengths. The observed degree sequence might be in principle replaced by its estimate from the first step of the method \cite{Cimini2015b}, 
yet this procedure becomes unfeasible for large networks, since it requires the solution of $2(N+M)$ coupled nonlinear equations. 
The alternative approach that we follow here is to make use of the CAPM framework, which can be seen as the bipartite counterpart of the standard gravity model 
used for monopartite networks~\cite{Wells2004,Mistrulli2011}. In its original version, CAPM assigns weights as 
\begin{equation}
\tilde\omega_{i\alpha}=V_i\,C_\alpha/W.
\label{eq:expect_w}
\end{equation}
As a result, the model reproduces the original strengths sequence only when fully connected networks are considered.
To account for a network topology determined by nontrivial connection probabilities, we can extend the dcGM prescription proposed in \cite{Cimini2015a} to bipartite networks. 
Weights of individual links are thus determined as:
\begin{equation}
\tilde{w}_{i\alpha}=\frac{V_i C_\alpha}{Wp_{i\alpha}}\tilde{a}_{i\alpha}=(z^{-1}+V_i\,C_\alpha)\,\frac{\tilde{a}_{i\alpha}}{W}.
\label{eq:weight}
\end{equation}
Notably, this equation ensures that the expected values of the reconstructed weights coincide with the CAPM ones, $\langle\tilde{w}_{i\alpha}\rangle_{\Omega}\equiv\tilde{\omega}_{i\alpha},\:\forall(i,\alpha)$, 
thus preserving the strength sequence on average, irrespectively from the underlying reconstructed topology~\cite{Cimini2015b}. 

\medskip

\add{\bf Remarks} --- To sum up, the set of connection probabilities defined in eq.~(\ref{eq:prob_plus}) and of link weights defined in eq.~(\ref{eq:weight}) univocally determine the ensemble $\Omega$, 
so that the value of the generic quantity $X$ in $G_0$ can be estimated as its ensemble average $\langle X \rangle_{\Omega}$.
\add{In the following, we will show results of the ECAPM reconstruction procedure for a case study of security holdings by European institutional sectors, 
which we believe can be of interest to both researchers and practitioners. A more exhaustive analysis of ECAPM can be found in the Appendix, 
where we show the analytic derivation of the model in the sparse network limit, as well as a more general formulation using the continuous approximation of the fitness model 
(in the latter case, we show how to write the expected values of the quantities of interest in terms of moments of the involved probability distributions).}

\medskip

Before moving further, let us explain the difference between the ECAPM and the MECAPM~\cite{DiGiangi2015} reconstruction techniques (see the Appendix for further details). 
Although both approaches predict average link weights matching the CAPM values, MECAPM leads to connection probabilities of the form $q_{i\alpha}=\tilde{\omega}_{i\alpha}/(1+\tilde{\omega}_{i\alpha})$, 
which is the special case $z=1/W$ of eq.~(\ref{eq:prob_plus}). These probabilities then assume values very close to unity even for moderately large weights. 
This is due to the very definition of MECAPM, obtained by maximizing entropy constraining only the strength sequence (\ie, without accounting for the network connectivity), 
a prescription which redistribute weights as evenly as possible---thus generating very dense networks.

\section{Data}

\begin{table}[t!]\footnotesize
 \begin{tabular}{l|l} 
 {\bf Sector} & {\bf Description} \\
 \hline
 s\_11 	 & Non-financial corporations \\
 s\_121	 & The central bank \\
 s\_122	 & Deposit-taking corporations except the central bank \\
 s\_123	 & Money market funds (MMFs) \\
 s\_124	 & Non-MMF investment funds \\
 s\_125	 & Other financial intermediaries, except ICPFs \\
 s\_126	 & Financial auxiliaries \\
 s\_127  & Captive financial institutions and money lenders \\ 
 s\_128  & Insurance corporations (ICs) \\
 s\_129	 & Pension funds (PFs) \\
 s\_13 	 & General government \\
 s\_14 	 & Households (only in holders) \\
 s\_15 	 & Non-profit insitutions serving households \\
 s\_X	 & Not allocated/unspecified \\
 \hline
\end{tabular}
\caption{Institutional sector classification by the European Central Bank.}\label{tab_sec}
\end{table}

\begin{figure*}[t!]
\begin{center}
\includegraphics[width=0.66\textwidth]{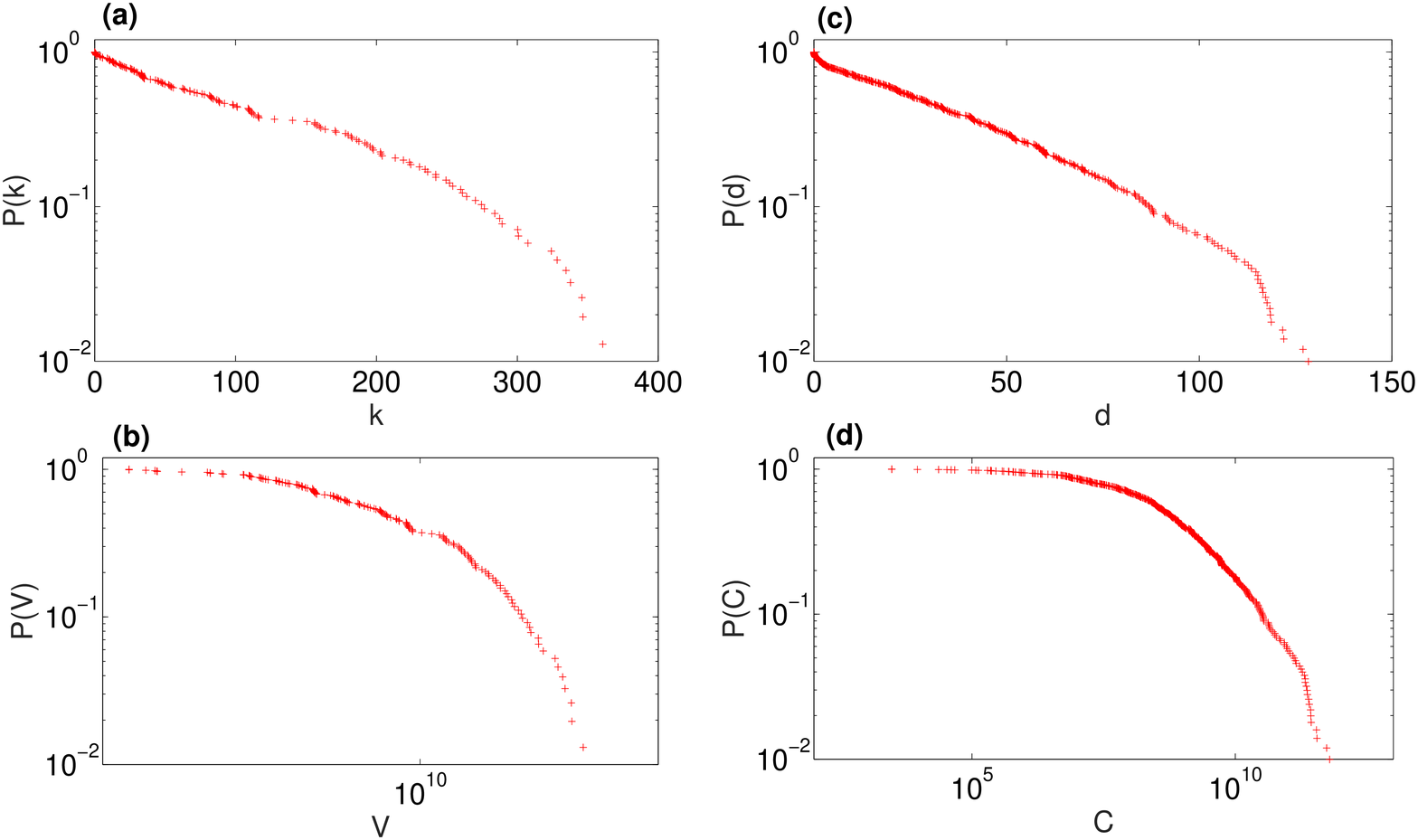}
\end{center}
\caption{Empirical cumulative distributions \add{for the degrees of holders $k$ (panel a) and of issuers $d$ (panel c), and of the strengths of holders $V$ (panel b) and of issuers $C$ (panel d),} 
in the SHS dataset.}
\label{fig-dist}
\end{figure*}

The dataset we use to test our reconstruction method is based on the {\em Security Holding Statistics} (SHS) collected by the European Central Bank. 
The data has quarterly information on each individual security at the country-sector level, covering a time span from $2009Q1$ to $2015Q2$, and is reported at the current euro value. 
Holding data include four types of financial tools: long-term security bonds, short-term security bonds, money market fund and equity. 
A distinctive feature of the dataset is that the holding information is collected for each International Securities Identification Number (ISIN). 

To carry out our analysis, we focus on long-term security bonds and consider quarterly snapshots of the debt exposure of 14 different institutional sectors, with standard ECB classification 
(see Table \ref{tab_sec}), for each of the 19 European reporting countries. This results in a first layer of $N=266$ holders nodes, each corresponding to a specific country-sector combination. 
The second layer represents the issuers of the securities, where we aggregated the specific ISINs to the same country-sector classification, with the exception of the households (only 13 sectors). 
Since the reporting countries are required to report on any equity, with no limitation on the location of security issuer, the second layer is composed of all the possible issuers nodes from 242 countries 
(including the 19 reporting countries) resulting in $M=3146$ country-sector combinations. 
Each link $w_{i\alpha}$ thus corresponds to the holdings of holder $i$ (aggregated to the country-sector) of the securities issued by issuer $j$ (aggregated to the country-sector).

Overall, we have 24 temporal snapshots (quarters) of the data, bearing the complex networks signature of broad degree and strength distributions (Fig. \ref{fig-dist}). 
Notice that dense reconstruction methods as CAPM and MECAPM would not be able to replicate these heterogeneous topological features.

We remark that that the magnitude of the collected holdings is substantial: 
for instance, total holdings by countries belonging to the Euro area amount to around 18.3 trillion euro at the end of June 2014, thus representing a relevant share of the total financial system. 
While the information on SHS marginals (\eg, the total asset of a particular holder sector) is relatively easy to access, the full availability of the SHS provides us with an unique opportunity 
to test the effectiveness of our method in reconstructing the network structure of such an important and large-scale financial system. 
For the sake of confidentiality, the analysis does not contain any explicit reference to any specific country. 
Also, for the sake of readability, in the following we report results only for the most recent quarter of the dataset $(2015Q2)$, yet our findings are robust across all 24 quarterly snapshots. 

\section{Results}\label{sec:test-bs}

\begin{figure}[t!]
\begin{center}
\includegraphics[width=0.45\textwidth]{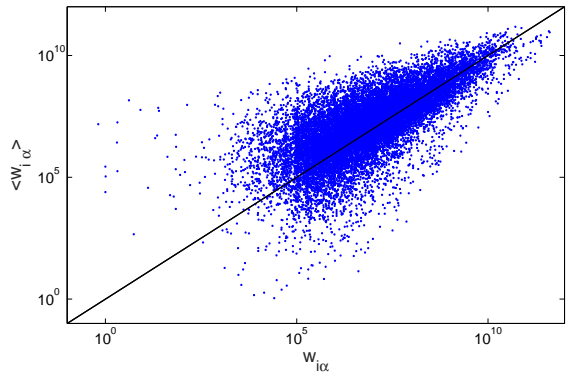}
\end{center}
\caption{Comparison between observed weights in the SHS network and CAPM-predicted ones. The black line denotes the identity.}\label{fig-weights}
\end{figure}

We start by looking at the size of individual exposures between holders and issuers, \ie, the weights of the existing links in the network. 
As already mentioned, ECAPM weight expectations coincide with those of the original CAPM and MECAPM defined in eq.~(\ref{eq:expect_w}). 
If we compare such expectations with the real observed weights (Fig. \ref{fig-weights}), we see that all reconstruction methods perform well in reproducing the non-zero weights of the network. 
However, weight uncertainties differ significantly between ECAPM and MECAPM, as shown by the ratio of standard deviations within the ensemble (see the Appendix):
\begin{equation}
r_{w_{i\alpha}}=\frac{\sigma_{w_{i\alpha}}^{\text{ECAPM}}}{\sigma_{w_{i\alpha}}^
{\text{MECAPM}}}\simeq\sqrt{\frac{1}{p_{i\alpha}}-1}.
\end{equation}
Since 90\% or the $r_{w_{i\alpha}}$ coefficients are smaller than 1 in our dataset, we can assert that ECAPM is more accurate than MECAPM. Notice the key role played by the topology 
in lowering the uncertainty of weights estimation: $r_{w_{i\alpha}}<1$ is equivalent to $p_{i\alpha}>1/2$. The fact that such a feature holds for country-sector holdings 
is also consistent with our fitness model ansatz on the relationship between node degrees and strengths.

\begin{figure*}[t!]
\begin{center}
\includegraphics[width=0.64\textwidth]{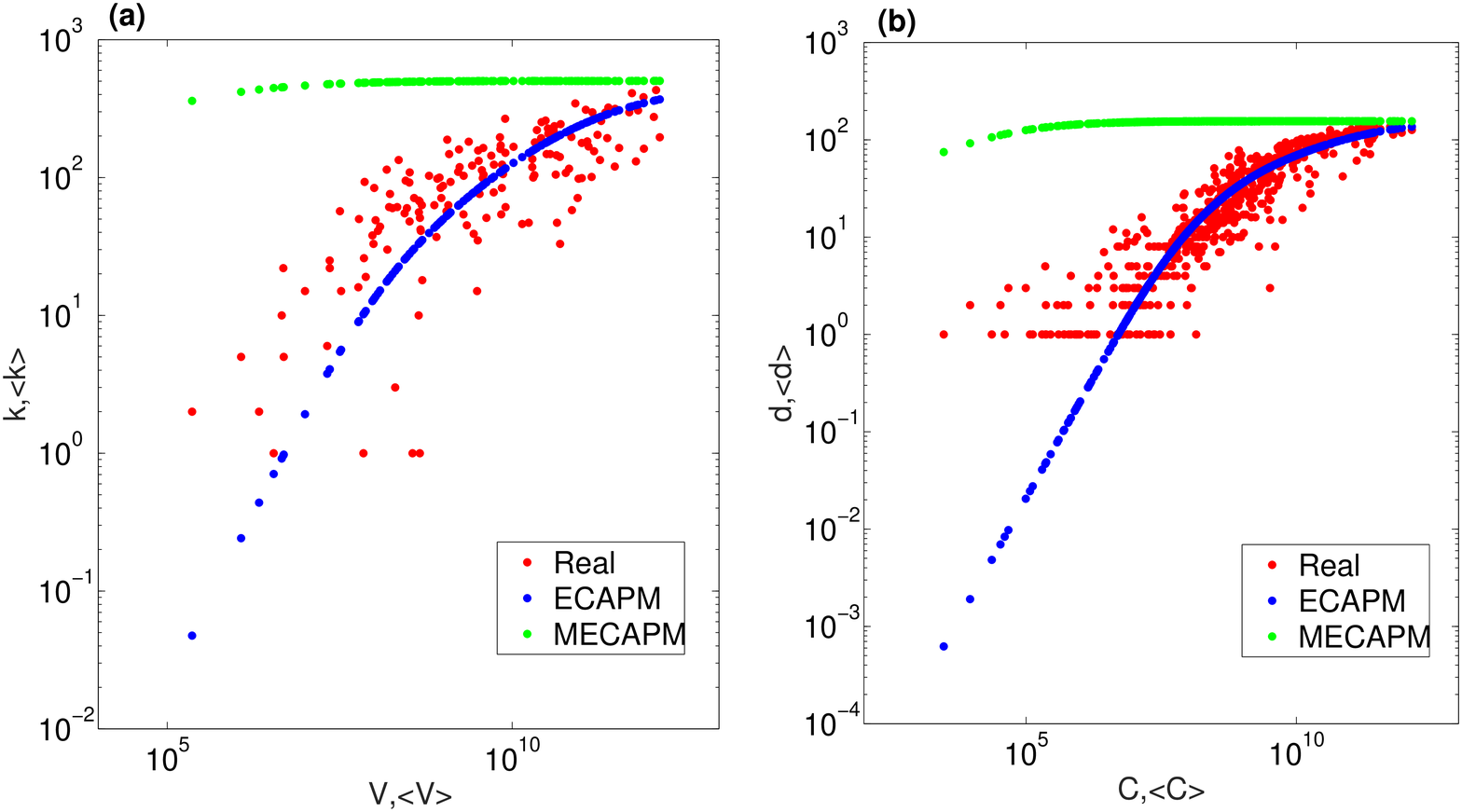}
\end{center}
\caption{Empirical relation between strengths and degrees in the data (red points), superimposed with the prediction by ECAPM (blue points) and MECAPM (green points), 
for both the \add{layers of holders (panel a) and of issuers (panel b).}}\label{fig-ansatz}
\end{figure*}

Next, we preform a key test for the fitness ansatz of our model (Fig. \ref{fig-ansatz}), 
showing that the strength-induced degrees $\{\langle k_i\rangle_{\Omega}\}_{i=1}^N$ and $\{\langle d_\alpha\rangle_{\Omega}\}_{\alpha=1}^M$ predicted by ECAPM 
interpolate well the empirical values $\{k_i(G_0)\}_{i=1}^N$ and $\{d_\alpha(G_0)\}_{\alpha=1}^M$, while MECAPM fails to a large extent. 
This result supports the use of connection probabilities defined by ECAPM as eq.~(\ref{eq:prob_plus}).
Then, to further assess the effectiveness of our reconstruction method, we consider three different families of indicators: topological, statistical and financial ones.

\begin{figure*}[t!]
\begin{center}
\includegraphics[width=0.7\textwidth]{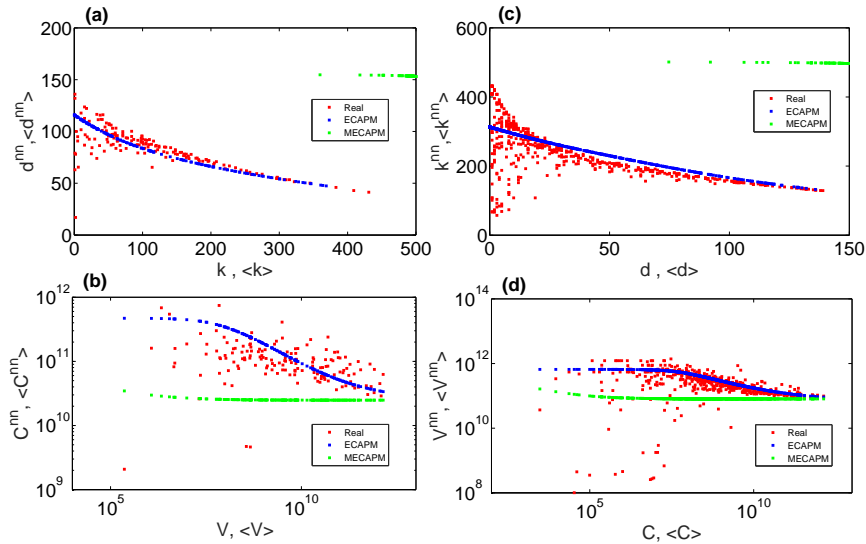}
\end{center}
\caption{Empirical values for the average nearest neighbors degrees $d^{nn}$ vs $k$ \add{(panel a)}, $k^{nn}$ vs $d$ \add{(panel c)} 
and the average nearest neighbors strengths $C^{nn}$ vs $V$ \add{(panel b)} and $V^{nn}$ vs $C$ \add{(panel d)} in the data (red points), 
superimposed with the prediction by ECAPM (blue points) and MECAPM (green points).}\label{fig:top_test}
\end{figure*}

Topological indicators aim at assessing whether a given reconstruction method can effectively reproduce higher-order features which are commonly regarded as most significant for describing a network.
Here we focus on the average nearest neighbors degrees $k^{nn}$ and $d^{nn}$ and strengths $V^{nn}$ and $C^{nn}$~\cite{Saracco2015}, 
namely the arithmetic mean of degrees and strengths of a node neighbors, respectively (see the Appendix). In the context of SHS data, $k^{nn}$ ($V^{nn}$) indicate how many exposures 
(the total portfolio weight) holders have on average for a specific issuer, while $d^{nn}$ ($C^{nn}$) indicate how many investors 
(the total amount of money invested) issuers have on average for a specific holder.
As Fig. \ref{fig:top_test} shows, our method correctly reproduces the observed decreasing trends, interpolating the clouds of points describing the real network. 
Again, the MECAPM predicts flat trends which are not compatible with the observed patterns, by systematically over/under-estimating them.

Statistical indicators instead provide information on the details of the reconstructed network structure. 
More specifically, they quantify the ability to correctly predict the presence of individual links (\ie, the position of ones in the binary biadjacency matrix) and their absence (\ie, the position of zeros). 
As explained in detail in the Appendix, this is measured by the performance of a binary classifier, 
thus using the usual percentages of {\it true positives}, {\it true negatives}, {\it false positives} and {\it false negatives}. 
Table \ref{tab1} reports four more informative combinations of such indices: {\it true positive rate} (TPR), {\it specificity} (SPC), {\it positive predicted value} (PPV) and {\it accuracy} (ACC). 
\add{Note that while these numbers refer to $2015Q2$, they are remarkably stable across all 24 quarters of our dataset (Fig. \ref{fig:fin_time}).} 
In order to correctly interpret the results provided in the table, we recall that while our ECAPM correctly replicates the observed link-density 
$c=\frac{L}{N\cdot M}\simeq 0.24$ (since $\langle L\rangle_{\Omega}=L$), 
MECAPM returns a density of $c_{\text{MECAPM}}\simeq 0.98$. As a trivial consequence, the MECAPM correctly places almost all the 1s, obtaining a very high true positive rate (TPR=0.99). 
However, this feature also represents a major drawback. In fact, the percentage of correctly recovered zeros is very low (SPC=0.02), 
pointing out a strong bias towards false positives (whose percentage amounts to FPR=1$-$SPC=0.98). 
Using the ROC curve, which illustrates the performance of a binary classifier as a point with coordinates (TPR,FPR) inside a unit square---the better the classifier, 
the closer its point to the (0,1) vertex---the point corresponding to MECAPM lies at the top right corner, close to the identity line: this, in turn, 
implies that its predictive power (PPV) equals that of a {\em random} classifier based on the bipartite random graph model 
(defined by the ``homogeneous'' assignments $p_{i\alpha}\equiv p=\frac{L}{N\cdot M}$ $\forall i,\alpha$). 
Thus the MECAPM performance in reproducing the SHS network structure is low, as confirmed by its small overall accuracy. In contrast, ECAPM performs much more accurately, 
correctly recovering $63\%$ of the ones and $90\%$ of zeros and thus achieving a much higher predictive power. 

\begin{table}[t!]
 \begin{tabular}{l|c|c} 
 					& {\bf ECAPM} & {\bf MECAPM} \\
 \hline
\text{True Positive Rate (TPR)}	&	0.63	&	0.99	\\
\text{Specificity (SPC)}	  	&	0.89	&	0.02	\\
\text{Positive Predicted Value (PPV)} &	0.63	&	0.24	\\
\text{Accuracy (ACC)}		&	0.83	&	0.25	\\ 
 \hline
\end{tabular}
\caption{Values of statistical indicators quantifying the performance of ECAPM and MECAPM in reproducing the details of the SHS network structure.}\label{tab1}
\end{table}

\begin{figure}[t!]
\begin{center}
\includegraphics[width=0.5\textwidth]{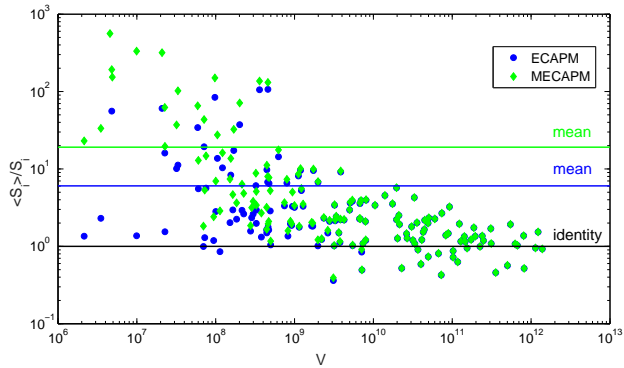}
\end{center}
\caption{Relative systemicness index $\tilde{S}_i/S_i(G_0)$ of the various country-sectors as estimated by ECAPM and MECAPM for a particular configuration drawn from the corresponding ensembles. 
Averages are shown for both methods as horizontal solid lines.}\label{fig:fin_test}
\end{figure}

Lastly, financial indicators provide information on how systemic risk is estimated in the reconstructed ensemble. 
\add{Since we are in the context of bipartite networks of portfolio holdings, here we focus on the risk stemming from sales of illiquid assets and consequent losses during fire sales. 
In particular,} we follow~\cite{Greenwood2015} and use the systemicness index $S_i$ as a measure of the impact of country-sector $i$ on the whole system:
\begin{equation}\label{eq.systemicness}
S_i=\frac{\Gamma_iV_i}{E}B_ir_i,
\end{equation}
where $\Gamma_i=\sum_{j\alpha}(w_{j\alpha}l_\alpha w_{i\alpha})$ \add{is computed from the illiquidity-weighted projection of the bipartite network}, $l_\alpha$ is the illiquidity parameter of $\alpha$, 
$B_i$ and $r_i$ are the leverage and portfolio return of country-sector $i$, and $E$ is the total equity in the system. 
\add{Note that systemicness as defined by eq.~(\ref{eq.systemicness}) is at the basis of more recent and refined risk measures \cite{Cont2016,Cont2017,Getmansky2017}, 
which fully implement dynamical downward spirals of asset prices in the system. Here, since our focus is on network reconstruction and not on systemic risk modeling, 
we refrain from using such metrics which are more difficult to handle. Yet, eq.~(\ref{eq.systemicness}) contains quantities which are not accessible through the SHS data. 
To get rid of them, we can assume homogeneous shocks in the system, as well as equal illiquidity for the asset classes, in order to define} the relative systemicness
\begin{equation}
\frac{\tilde S_i}{S_i(G_0)}=\frac{\sum_{j\alpha}\tilde{w}_{j\alpha}\tilde{w}_{i\alpha}}{\sum_{j\alpha}w_{j\alpha}w_{i\alpha}}.
\end{equation}
We can then use this quantity to test the reconstruction procedures. After some algebra (reported in Appendix), we find that ECAPM and MECAPM lead to the same predictions 
in terms of ensemble average for the systemicness. This is a natural consequence of weights expectations which coincide in the two algorithms. 
However, when single instances of networks drawn from the corresponding reconstructed ensembles are considered, the two methods again differ. 
Fig. \ref{fig:fin_test} shows the relative systemicness values scattered versus the holdings of country-sectors. 
As for the reconstruction of individual weights, the estimates provided by the two methods coincide for the largest nodes. However, MECAPM tends to overestimate the systemicness of small nodes. 
More importantly, the two methods perform quite differently in estimating the standard deviation of systemicness values. In particular, 
the standard deviations of the values shown in Fig. \ref{fig:fin_test} are $\varsigma^{\text{MECAPM}}\simeq 2\cdot 10^4$ and $\varsigma^{\text{ECAPM}}\simeq 10$, 
and the ratio of the standard deviations within the ensemble, $r_{S_i}=\sigma_{S_i}^{\text{ECAPM}}/\sigma_{S_i}^{\text{MECAPM}}$, is smaller than 1 for 90\% of the nodes in the network.
\add{This is particularly relevant in potential stress test applications, since the range of variability of the index within the reconstructed ensemble basically sets the scenarios to be considered: 
having a small variability is desirable to focus on the most plausible network configurations.}

\section{Discussion}\label{sec:conclusions} 

Reconstructing the properties of a network when only partial information is available represents a key open problem in the field of complex systems. 
Our proposed solution points out that the accuracy of a reconstruction method crucially depends on its ability to reproduce the network topology (\ie, the presence of single links).
In fact, without enforcing any topological constraints (\eg, whenever only the strength sequence is enforced), the number of connections is usually overestimated, 
causing an excess of false positives and, in turn, reducing the predictive power of the reconstruction method. 
Indeed, a method constraining only nodes strengths assigns the same probability to all configurations with the same strength sequence: 
since, as pointed out in previous contributions~\cite{Mastrandrea2014,Cimini2015b}, the number of configurations satisfying these constraints is usually extremely large, 
the accuracy of the achieved reconstruction is undermined.

The fundamental role played by topology clearly emerges in the calculation of the errors affecting the estimates of quantities considered. 
In our case, both the standard deviations for the weights and of the systemicness index are defined in terms of linking probabilities. 
Remarkably, lower errors are obtained in correspondence of higher probabilities: according to our assumption, this implies that bigger nodes establish more connections, 
which are in turn less affected by errors. This is relevant especially in terms of systemic risk: bigger nodes are also the most dangerous ones in case of severe losses or default. 
However, although errors are sensitive to topological details, this is much less the case for the systemicness index measure. 
$S$, in fact, ignores the node-specific patterns of connections, being purely defined in terms of links weights (see the Appendix). 
As a consequence, any algorithm providing a satisfactory estimate of the latter performs equally well, irrespectively of its performance in reproducing the purely structural quantities 
(as the presence of the links themselves). In our view, this can represent a drawback of the $S$ index to measure systemic risk.
Other measures, on the other hand, are sensitive to topology~\cite{Caccioli2014,Corsi2013}. In particular, in~\cite{Caccioli2014} it is shown that the probability of contagion in stock markets 
strongly depends on connectivity, achieving the minimum value for both the empty and the fully connected configurations and the maximum value for intermediate values of the link density 
(while the extent of the contagion is a monotonic function of the density). In~\cite{Corsi2013}, on the other hand, it is pointed out that portfolio heterogeneity affects the extent 
and the coordination of feedback effects ``triggering transition from stationary dynamics of price returns to non-stationary ones characterized by a steep growth (bubbles) and plunges (bursts) of market prices''. 
Indeed, the analysis of historical data sets that also encompass bubbles and crises represents the subject of future work.

\add{By relying on the fitness model ansatz, ECAPM defines a null model for bipartite weighted networks where strengths replace degree constraints. 
When available, degrees can be directly used to define the topology using the full BiCM approach \cite{Saracco2015}. 
If this is not the case,} the applicability of the method strongly depends on the accuracy of the assumed functional relationship between strengths and degrees. 
While the validity of the fitness ansatz can be assessed by explicitly comparing nodes strengths and degrees (whenever available), its rationale rests upon a simple argument, 
corroborated by the analysis of other economic and financial networks \cite{Garlaschelli2004,DeMasi2006,Garlaschelli2005,Cimini2015a,Cimini2015b}: 
the higher the importance of a node, the larger we expect its degree to be. \add{Indeed, our method can be applied to any weighted bipartite network for which this argument applies.}
Similarly, the factorization of any graph probability into the product of single link probabilities characterizing the ECAPM model could be perceived 
as inherently inadequate to reproduce the real network patterns. Although this may indeed be the cause of the residual deviations between the actual network values and the reconstructed ones, 
we emphasize that the independence of link probabilities is not postulated by us at any stage: rather, it emerges naturally from the enforcement of purely local constraints (be they degrees or strengths). 
Put differently, the unbiased solution to the inference problem prescribes to consider independent probabilities. 
Paradoxically, introducing (more realistic) dependencies would result in a biased inference, if starting from the same observational constraints.

\section{Acknowledgements}
We thank Stefano Battiston for useful discussions. 
This work was supported by the EU projects MULTIPLEX (grant num. 317532), DOLFINS (grant num. 640772), CoeGSS (grant num. 676547), Shakermaker (grant num. 687941), SoBigData (grant num. 654024). 
DG acknowledges support from the Econophysics foundation (Stichting Econophysics, Leiden, the Netherlands).

\section{Author Contribution}
G Caldarelli, G Cimini, D Garlaschelli and T Squartini developed the method. A Almog, G Cimini and T Squartini designed and performed the analysis. I van Lelyveld provided the data set. 
G Cimini and T Squartini wrote the manuscript. All authors reviewed and approved the manuscript.

\setcounter{table}{0}
\setcounter{figure}{0}
\setcounter{equation}{0}
\renewcommand{\thetable}{A\arabic{table}}
\renewcommand{\thefigure}{A\arabic{figure}}
\renewcommand{\theequation}{A\arabic{equation}}
\renewcommand{\theHtable}{A\arabic{table}}
\renewcommand{\theHfigure}{A\arabic{figure}}
\renewcommand{\theHequation}{A\arabic{equation}}

\section*{Appendix}

\subsection*{The MECAPM method}

MECAPM ({\it maximum-entropy capital-asset pricing model}) combines the 
constrained entropy-maximization (ME) technique and the single weights 
estimation $w_{i\alpha}$ from the CAPM model~\cite{DiGiangi2015}, which reads
\begin{equation}
\langle 
w_{i\alpha}\rangle_{\text{CAPM}}=\frac{V_iC_\alpha}{W}\qquad\forall(i,\alpha).
\end{equation}
This is achieved using the exponential random graph theoretical framework~\cite{Park2004,Squartini2011}. 
Indeed, the CAPM model can be understood as prescribing to constrain the whole set of weights of 
a given weighted undirected bipartite network $G_0$. In this particular 
case, the probability distribution over the ensemble factorizes into the 
product of pair-specific geometric probability distributions
\begin{equation}
P(G_0)=\prod_{i}\prod_{\alpha}q_{i\alpha}^{w_{i\alpha}}(1-q_{i\alpha}).
\end{equation}
The generic weight $w_{i\alpha}$ is interpreted as the sum of unitary links, 
each one behaving as an independent Bernoulli random variable. Each pair-specific probability distribution is thus characterized by a 
quantity $q_{i\alpha}$, describing the probability that nodes $i$ and $\alpha$ are connected.

The $N\cdot M$ unknown parameters $q_{i\alpha}$ ($i=1\dots N$, $\alpha=1\dots 
M$) are estimated by exploiting the likelihood conditions 
\cite{Squartini2011}
\begin{equation}
\langle 
w_{i\alpha}\rangle_{\text{CAPM}}=\frac{q_{i\alpha}}{1-q_{i\alpha}}
\qquad\forall(i,\alpha);
\end{equation}
as a result, one obtains 
\begin{equation}\label{SI1}
q_{i\alpha}=\frac{\langle w_{i\alpha}\rangle_\text{CAPM}}{1+\langle 
w_{i\alpha}\rangle_\text{CAPM}}=\frac{V_iC_\alpha/W}{1+V_iC_\alpha/W}
\qquad\forall(i,\alpha).
\end{equation}

\add{Note that the probability coefficients described by eq. (\ref{SI1}) solely depend on the (estimated) weights magnitude: 
as a consequence, the larger the value $\langle w_{i\alpha}\rangle$, the larger the probability that nodes $i$ and $\alpha$ are connected. 
Since weights are typically large, the MECAPM method is likely to predict very dense configurations.} 
For what concerns the SHS dataset considered in the present work, although link probabilities range in the interval $[5\cdot 10^{-5},\:1]$, 
their distribution is characterized by an average value of $\overline{q}\simeq0.97$ and a standard deviation of $\sigma_q\simeq 0.091$. 
Their variation coefficient is $\sigma_q/\overline{q}\simeq 0.093$, indicating that the average faithfully represents the distribution of values (which, practically, coincide with a delta peaked at 1).

\subsection*{The ECAPM method}

Differently from MECAPM, the ECAPM algorithm features a correction term $1/z$ that accounts for the overall network link density. 
For the SHS dataset, probability coefficients computed by ECAPM range in the interval $[10^{-12},\:1]$, their average value is $\overline{p}\simeq0.01$ (coinciding with the network density of links) 
and their standard deviation is $\sigma_p\simeq 0.057$. Their variation coefficient is $\sigma_p/\overline{p}>5$, indicating that the average is not representative of the distribution of values, 
which is thus rather spread over the corresponding support.

\add{The key ingredient of ECAPM is to quantify the tendency of any two nodes of establishing a connection, by assuming the former to depend on the nodes fitness and on the (observed) total number of links. 
The way in which the ECAPM method works can be better illustrated by considering two special cases which are analytically tractable.}

\add{\paragraph{Sparse networks.} When considering low-density networks, the probability coefficients can be assumed to be ``small''. 
Let us, thus, assume that the Taylor expansion (around $z_0=0$) of any coefficient can be truncated to the first order:
\begin{eqnarray}\label{tay}
p_{i\alpha}=\frac{zV_iC_\alpha}{1+zV_iC_\alpha}&=&zV_iC_\alpha-(zV_iC_\alpha)^2+(zV_iC_\alpha)^3\dots\nonumber\\
&\simeq& zV_iC_\alpha.
\end{eqnarray}
In this case, $z$ can be estimated by imposing $\langle L\rangle=\sum_i\sum_\alpha p_{i\alpha}=L$, which leads to $z=L/W^2$, further implying $p_{i\alpha}=L(V_iC_\alpha/W^2)$. 
The latter expression highlights that any two nodes are connected with a probability that is a fraction of the total number of links, 
properly normalized by the percentage of the total weight represented by the two involved nodes strengths. 
Analogously, when calculating quantities like nodes degrees, one finds that $\langle k_i\rangle=\sum_\alpha p_{i\alpha}=V_i(L/W)$ and $\langle d_\alpha\rangle=\sum_i p_{i\alpha}=C_\alpha(L/W)$.}

\add{\paragraph{The continuous approximation.} In the sparse case, the empirical distributions of strengths are directly proportional to the empirical distributions of degrees. 
As a consequence, a scale-free distribution of the former induces a scale-free distribution of the latter. Such an evidence can be used to gain insight into ECAPM also in the general case: 
upon assuming nodes fitnesses (\ie, nodes strengths) to be drawn from well-defined probability distributions, $V\sim\pi(V)$ and $C\sim\rho(C)$, 
it is enough to replace each sum with an integral over the aforementioned distributions. 
As an example, the formula for computing the degree of portfolio $i$, $\langle  k_i\rangle=\sum_\alpha p_{i\alpha}$, becomes
\begin{equation}\label{cont}
\langle k_i\rangle(z, V)=M\int_{\mathcal{C}}\left(\frac{zV_iC}{1+zV_iC}\right)\rho(C)dC
\end{equation}
with $\mathcal{C}$ representing the support of the distribution. By inserting the Taylor expansion in eq. (\ref{tay}) into eq. (\ref{cont}) we get:
\begin{eqnarray}\label{cont2}
\langle k_i\rangle&(z, V_i)&\,=\,M\int_{\mathcal{C}}\left(\frac{zV_iC}{1+zV_iC}\right)\rho(C)dC=\nonumber\\
&=&M\int_{\mathcal{C}}[zV_iC-(zV_iC)^2+(zV_iC)^3\dots]\rho(C)dC=\nonumber\\
&=&M[zV_i\mu_1-(zV_i)^2\mu_2+(zV_i)^3\mu_3\dots],
\end{eqnarray}
offering a recipe for calculating the numerical value of the statistics of interest, via the computation of the (raw) moments of the involved distributions. 
As a last comment, note that the formulas for the sparse case is obtained by retaining only the first term in eq. (\ref{cont2}), and requiring that the first moment of $\rho(C)$ 
can be estimated by invoking the likelihood-maximization principle, \ie, $\mu=\overline{C}=(\sum_\alpha C_\alpha)/M=W/M$.}

\medskip

\subsection*{Weights estimation}

As stated in the main text, the ensemble averages of weights under the MECAPM and ECAPM coincide. However, the corresponding statistical fluctuations differ:
\begin{eqnarray}
\sigma^{2,\:\text{ECAPM}}_{w_{i\alpha}}&=&\langle 
w_{i\alpha}\rangle_{\text{CAPM}}^2\left[\frac{1}{p_{i\alpha}}-1\right],\\
\sigma^
{2,\:\text{MECAPM}}_{w_{i\alpha}}&=&\langle 
w_{i\alpha}\rangle_{\text{CAPM}}(1+\langle w_{i\alpha}\rangle_{\text{CAPM}}).
\end{eqnarray}
This implies
\begin{eqnarray}
r_{w_{i\alpha}}&=&\frac{\sigma_{w_{i\alpha}}^\text{ECAPM}}{\sigma_{w_{i\alpha}}
^\text{MECAPM}}=\nonumber\\
&=&\frac{\langle w_{i\alpha}\rangle_{\text{CAPM}}^2}{\langle 
w_{i\alpha}\rangle_{\text{CAPM}}(1+\langle 
w_{i\alpha}\rangle_{\text{CAPM}})}\sqrt{\frac{1}{p_{i\alpha}}-1}=\nonumber\\
&\simeq&\sqrt{\frac{1}{p_{i\alpha}}-1},
\end{eqnarray}
a ratio that is (strictly) smaller than 1 whenever $p_{i\alpha}>1/2$.

\subsection*{Testing the reconstruction methods}

In order to test the performance of both ECAPM and MECAPM in reconstructing the network, we have considered {\it topological}, {\it statistical} and {\it financial} indicators.

\subsubsection*{Topological indicators}

The first family of indicators consists of 
quantities providing information on the global, structural organization of a 
given network. For what concerns the network binary structure, we have 
considered i) the {\it degree} of, respectively, holders and issuers, 
$k_i(\mathbf{A})=\sum_{\alpha=1}^M a_{i\alpha}$ and 
$d_\alpha(\mathbf{A})=\sum_{i=1}^Na_{i\alpha}$ and ii) the {\it average nearest 
neighbors degree} of holders and issuers \cite{Saracco2015}:
\begin{equation}
d_i^{nn}(\mathbf{A})=\frac{\sum_{\alpha=1}^Ma_{i\alpha}d_{\alpha}}{k_i},\;k_\alpha^{nn}(\mathbf{A})=\frac{\sum_{i=1} ^Na_{i\alpha}k_i}{d_\alpha}
\end{equation}
We compared these values with their expected counterparts, \ie, $\langle k_i\rangle=\sum_{\alpha=1}^M \langle a_{i\alpha}\rangle$, $\langle d_\alpha\rangle=\sum_{i=1}^N \langle a_{i\alpha}\rangle$ and
\begin{equation}
\langle d_i^{nn}\rangle=\frac{\sum_{\alpha=1}^M \langle a_{i\alpha}\rangle\langle d_{\alpha}\rangle}{\langle k_i\rangle},\:
\langle k_\alpha^{nn}\rangle=\frac{\sum_{i=1}^N \langle a_{i\alpha}\rangle\langle k_i\rangle}{\langle d_\alpha\rangle}.
\end{equation}
\add{Note that, for ECAPM in the sparse case, it is $\langle d_i^{nn}\rangle\simeq (L/W^2)\sum_\alpha C_\alpha^2$ and $\langle k_\alpha^{nn}\rangle\simeq (L/W^2)\sum_i V_i^2$, 
\ie, the expected value of the average nearest neighbors degree of holders and issuers is constant for all nodes belonging to the same layer.}
The predictions by MECAPM, on the other hand, can be obtained by considering a very dense network, \ie, $\langle a_{i\alpha}\rangle=q_{i\alpha}\simeq 1$. 
In this case, we obtain $\langle k_\alpha^{nn}\rangle_{\text{MECAPM}}\simeq M-1$ and $\langle d_i^{nn}\rangle_{\text{MECAPM}}\simeq N-1$.

\medskip

For what concerns the weighted structure of the network, we have considered the {\it average nearest neighbors strengths}
\begin{equation}
C_i^{nn}(\mathbf{W})=\frac{\sum_{\alpha=1}^M a_{i\alpha}C_{\alpha}}{k_i},\;V_\alpha^{nn}(\mathbf{W})=\frac{\sum_{i=1} ^N a_{i\alpha}V_i}{d_\alpha}
\end{equation}
and compared them with their expected counterparts
\begin{equation}
\langle C_i^{nn}\rangle=\frac{\sum_{\alpha=1}^M \langle a_{i\alpha}\rangle C_{\alpha}}{\langle k_i\rangle},\:
\langle V_\alpha^{nn}\rangle=\frac{\sum_{i=1} ^N \langle a_{i\alpha}\rangle V_i}{\langle d_\alpha\rangle}.
\end{equation}
\add{As for the average nearest neighbors degrees, in the sparse case ECAPM predicts flat trends for all nodes belonging to the same layer: 
$\langle C_i^{nn}\rangle\simeq (\sum_\alpha C_\alpha^2)/W$ and $\langle k_\alpha^{nn}\rangle\simeq (\sum_i V_i^2)/W$.}
MECAPM predictions are instead obtained by again considering a very dense network: $\langle V_\alpha^{nn}\rangle_{\text{MECAPM}}\simeq W/(N-1)$, $\langle C_i^{nn}\rangle_{\text{MECAPM}}\simeq W/(M-1)$.

\subsubsection*{Statistical indicators}

Statistical indicators are compactly represented 
by the so-called {\it confusion matrix}, a $4\times4$ matrix whose 
entries represent the number of {\it true positives}, {\it true negatives}, {\it 
false positives} and {\it false negatives}. Let us briefly explain why these 
concepts are useful for our analysis.

Reconstructing a network $G_0$ means providing an algorithm to estimate 
the {\it presence} and the {\it weight} of the connections. If we limit our 
analysis to the binary structure only (represented by the binary matrix 
$\mathbf{A}$, with $a_{i\alpha}=\Theta[w_{i\alpha}]$), this implies ``guessing'' 
the topological structure of the network, namely the position of 0s and 1s in the matrix. 
For each entry of the considered biadjacency matrix, four different cases are 
possible: {\bf a)} $a_{i\alpha}=1$ and we correctly predict 
$\tilde{a}_{i\alpha}=1$. When this is the case, we have a {\it true positive}; 
{\bf b)} $a_{i\alpha}=1$ but we predict $\tilde{a}_{i\alpha}=0$. In this case, 
we have a {\it false negative}; {\bf c)} $a_{i\alpha}=0$ and we correctly 
predict $\tilde{a}_{i\alpha}=0$. In this case, we have a {\it true negative}; 
{\bf d)} $a_{i\alpha}=0$ but we predict $\tilde{a}_{i\alpha}=1$. When this is 
the case, we have a {\it false positive}. 

Given the observed biadjacency matrix $\mathbf{A}$ and the reconstructed matrix 
$\mathbf{\tilde{A}}$, we can straightforwardly count the total number of true 
positives as the point-wise product of the two matrices
\begin{equation}
TP=\sum_{i}\sum_{\alpha}a_{i\alpha}\tilde{a}_{i\alpha}
\end{equation}
(whose generic addendum is one iff both $a_{ij}=1$ and $\tilde{a}_{ij}=1$), the 
total number of false negatives as
\begin{equation}
FN=\sum_{i}\sum_\alpha a_{i\alpha}(1-\tilde{a}_{i\alpha})=L(G_0)-TP
\end{equation}
(where $L(G_0)$ is total number of observed links), the total number of true 
negatives as
\begin{equation}
TN=\sum_i\sum_{\alpha}(1-a_{i\alpha})(1-\tilde{a}_{i\alpha})=N\cdot 
M-\tilde{L}-L(G_0)+TP
\end{equation}
(whose generic addendum is one iff both $a_{ij}=0$ and $\tilde{a}_{ij}=0$), and 
the total number of false positives as
\begin{equation}
FP=\sum_i\sum_\alpha(1-a_{i\alpha})\tilde{a}_{i\alpha}=\tilde{L}-TP=(N\cdot 
M-L(G_0))-TN
\end{equation}
whose first addendum is the number of 0s in the real matrix $\mathbf{A}$.

The information provided by $TP$, $FN$, $TN$, $FP$ is usually given by combinations of them. 
{\it Sensitivity} (or {\it true positive rate}) is defined as
\begin{equation}
TPR=\frac{TP}{TP+FN}=\frac{TP}{L(G_0)}
\end{equation}
and quantify the percentage of 1s that are correctly recovered by a reconstruction method. {\it Specificity} (or {\it true negative rate}) is defined as
\begin{equation}
SPC=\frac{TN}{FP+TN}=\frac{TN}{N\cdot M-L(G_0)}
\end{equation}
and quantifies the percentage of 0s that are correctly recovered by a reconstruction method. The {\it false positive rate}, defined as $FPR=1-SPC$, is usually compared to TPR 
in order to evaluate the performance of a given algorithm with respect to a random classifier. In fact, whenever TPR and FPR coincide, the performance of the considered model 
equals that of a random classifier (in other words, the model representative point lies on the identity line of the ROC curve). {\it Precision} (or {\it positive predicted value}) is defined as
\begin{equation}
PPV=\frac{TP}{TP+FP}
\end{equation}
and measures the performance of an algorithm in correctly placing the 1s with respect to the total number of predicted 1s. Finally, {\it accuracy} is
\begin{equation}
ACC=\frac{TP+TN}{TP+TN+FP+FN}=\frac{TP+TN}{N\cdot M}
\end{equation}
and measures the overall performance of a reconstruction method in correctly placing both 1s and 0s.

\medskip

\add{Since both the ECAPM and the MECAPM methods deal with an entire ensemble of candidate matrices $\mathbf{\tilde{A}}$, we are interested in estimating the {\em expected} values of the aforementioned indices. 
Their ensemble averages are reported in Table \ref{tabA1}. For the sake of illustration, let us explicitly derive them for the ECAPM method in the sparse regime and for the MECAPM method, below.}

\add{By resting upon the continuous approximation, the expected number of true positives for ECAPM is
\begin{eqnarray}\label{tpr}
\langle TP\rangle&=&L(G_0)\int_{\mathcal{V}}\int_{\mathcal{C}}p(V, C)\rho(C)\pi(V)dCdV=\nonumber\\
&\simeq&L(G_0)\int_{\mathcal{V}}\int_{\mathcal{C}}(zVC)\rho(C)\pi(V)dCdV=\nonumber\\
&=&L(G_0)z\lambda\mu=L(G_0)^2/(N\cdot M),
\end{eqnarray}
where we have truncated the analytical expression of $p(V, C)$ at the first order and estimated the first raw moment of the two distributions as $\lambda=\overline{V}=W/N$ and $\mu=\overline{C}=W/M$. 
In the same regime of eq. (\ref{tpr}), $\langle TPR\rangle=L(G_0)/(N\cdot M)$, \ie, the expected sensitivity coincides with the link density.}

\add{In the case of MECAPM, instead, TPR and FPR practically coincide (see Table \ref{tab1}).} This implies
\begin{eqnarray}
\langle TPR\rangle&=&1-\langle SPC\rangle\\
\Rightarrow\frac{\langle TP\rangle}{L_{\text{true}}}&=&\frac{\langle L\rangle-\langle TP\rangle}{N\cdot M-L_{\text{true}}}\nonumber\\
\Rightarrow\frac{\langle TP\rangle}{\langle L\rangle}&=&\langle PPV\rangle=\frac{L(G_0)}{N\cdot M},\nonumber
\end{eqnarray}
\ie, the MECAPM predictive power coincides with the network link density, which is the PPV of the random graph model. 

\add{Interestingly, the performance of ECAPM method in the sparse case and that of MECAPM in recovering the correct number of 1s are very similar (as confirmed by their close PPV values). 
However, in the regime of eq. (\ref{tpr}), the expected accuracy for ECAPM reads $\langle ACC\rangle\simeq 1+2L(G_0)/(N\cdot M)$, which can be quite large if the network is very sparse. 
In other words, the overall accuracy of ECAPM in reconstructing sparse networks can still be large, the reason lying in the large number of 0s correctly recovered.}

\begin{table*}[t!]
\[\begin{array}{c|c}
{\bf ECAPM} & {\bf MECAPM} \\
\hline\\ [1ex]
\langle TP\rangle_{\text{ECAPM}}=\sum_{i}\sum_{\alpha}a_{i\alpha}p_{i\alpha} & 
\langle TP\rangle_{\text{MECAPM}}\simeq L(G_0)\\ [2ex]
\hline\\ [1ex]
\langle 
TN\rangle_{\text{ECAPM}}=\sum_i\sum_{\alpha}(1-a_{i\alpha})(1-p_{i\alpha}
)=N\cdot M-2L(G_0)+\langle TP\rangle & \langle 
TN\rangle_{\text{MECAPM}}\simeq0\\ [2ex]
\hline\\ [1ex]
\langle FP\rangle_{\text{ECAPM}}=(N\cdot M-L(G_0))-\langle 
TN\rangle_{\text{ECAPM}}=L(G_0)-\langle TP\rangle_{\text{ECAPM}} & 
\langle FP\rangle_{\text{MECAPM}}\simeq N\cdot M-L(G_0)\\ [2ex]
\hline\\ [1ex]
\langle FN\rangle_{\text{ECAPM}}=L(G_0)-\langle 
TP\rangle_{\text{ECAPM}} & \langle FN\rangle_{\text{MECAPM}}\simeq0\\ [2ex]
\hline
\hline\\ [1ex]
\langle TPR\rangle_{\text{ECAPM}}=\langle 
TP\rangle_{\text{ECAPM}}/L(G_0) & \langle 
TPR\rangle_{\text{MECAPM}}\simeq1\\ [2ex]
\hline\\ [1ex]
\langle FPR\rangle_{\text{ECAPM}}=\frac{L(G_0)-\langle 
TP\rangle_{\text{ECAPM}}}{N\cdot M-L(G_0)} & \langle 
FPR\rangle_{\text{MECAPM}}\simeq1\\ [2ex]
\hline\\ [1ex]
\langle SPC\rangle_{\text{ECAPM}}=\frac{\langle TN\rangle_{\text{ECAPM}}}{N\cdot 
M-L(G_0)}=1-\langle FPR\rangle_{\text{ECAPM}} & \langle 
SPC\rangle_{\text{MECAPM}}\simeq0\\ [2ex]
\hline\\ [1ex]
\langle PPV\rangle_{\text{ECAPM}}=\frac{\langle 
TP\rangle_{\text{ECAPM}}}{\langle TP\rangle_{\text{ECAPM}}+\langle 
FP\rangle_{\text{ECAPM}}}=\frac{\langle 
TP\rangle_{\text{ECAPM}}}{L(G_0)} & \langle 
PPV\rangle_{\text{MECAPM}}\simeq\frac{L(G_0)}{N\cdot 
M}=c(G_0)\\ [2ex]
\hline\\ [1ex]
\langle ACC\rangle_{\text{ECAPM}}=\frac{\langle TP\rangle_{\text{ECAPM}}+\langle 
TN\rangle_{\text{ECAPM}}}{N\cdot M}=1-2c(G_0)+2\frac{\langle 
TP\rangle_{\text{ECAPM}}}{N\cdot M} & \langle ACC\rangle_{\text{MECAPM}}\simeq 
c(G_0)\\ [2ex]
\hline
\end{array}\]
\caption{Statistical indicators used to evaluate the performance of ECAPM and MECAPM in reproducing the observed network structure.}\label{tabA1}
\end{table*}

\begin{figure*}[t!]
\begin{center}
\includegraphics[width=0.66\textwidth]{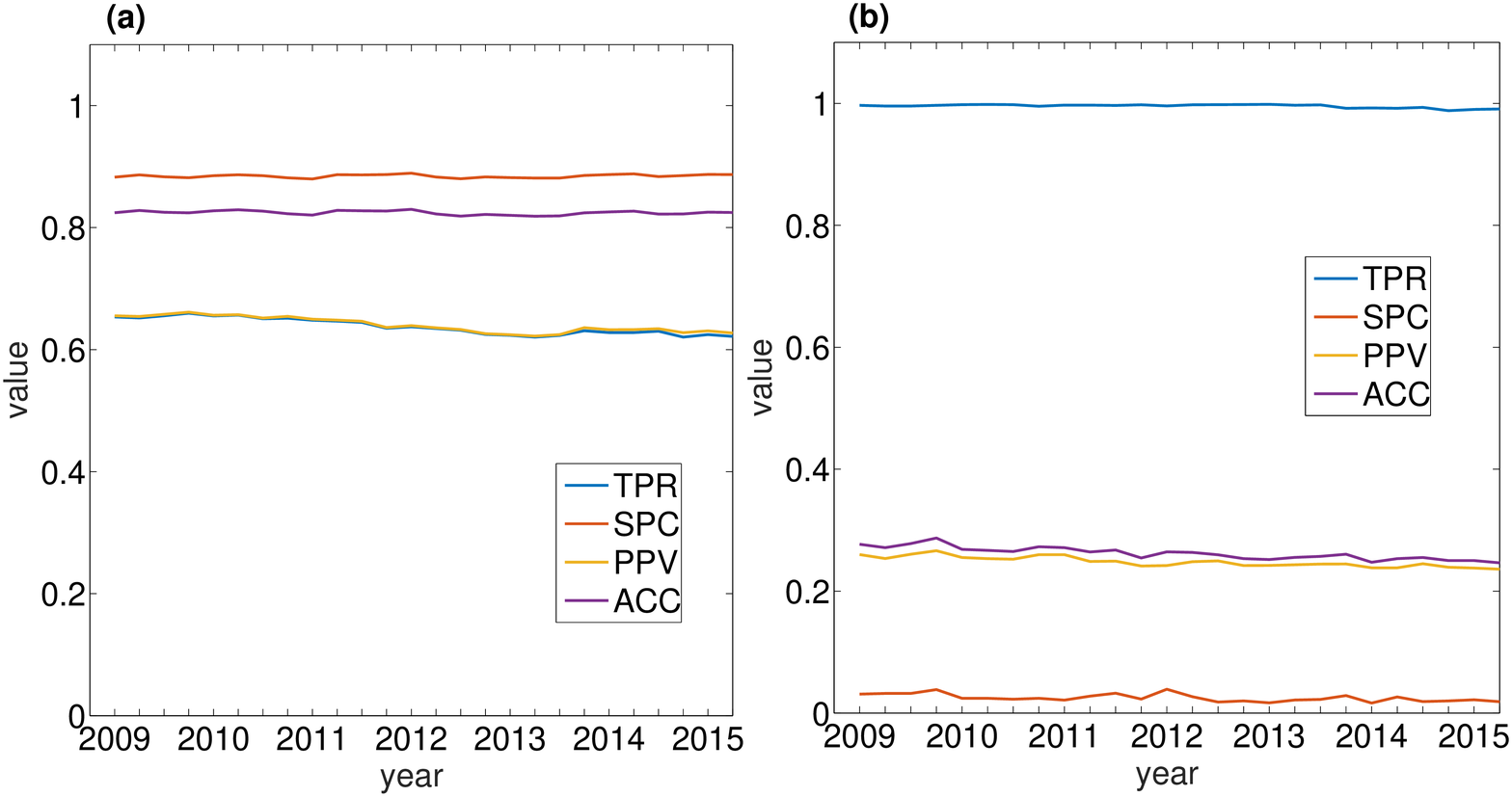}
\end{center}
\caption{Performance of ECAPM \add{(panel a)} and MECAPM \add{(panel b)} measured by statistical indicators across the time span of the SHS dataset.}\label{fig:fin_time}
\end{figure*}

\subsubsection*{Financial indicators}

The third family of indicators aim at quantifying the systemicness of nodes belonging to a financial 
network. We follow \cite{Greenwood2015} and adopt the systemicness index $S_i$:
\begin{equation}
S_i=\frac{\Gamma_iV_i}{E}B_ir_i
\end{equation}
which can be compared with its expected counterpart $\langle S_i\rangle$ via the 
ratio $\langle S_i\rangle/S_i$. After some algebraic manipulations 
\cite{DiGiangi2015}, such a ratio can be rewritten as
\begin{equation}
\frac{\langle S_i\rangle}{S_i}=\frac{\sum_j\sum_\alpha \langle 
w_{i\alpha}w_{j\alpha}\rangle}{\sum_j\sum_\alpha 
w_{i\alpha}w_{j\alpha}}=\frac{\sum_\alpha\langle 
w_{i\alpha}^2\rangle+\sum_{j\neq i}\langle w_{i\alpha}\rangle\langle 
w_{j\alpha}\rangle}{\sum_\alpha w_{i\alpha}C_\alpha}.
\end{equation}

For ECAPM, this ratio can be further simplified upon 
considering that $\langle w_{i\alpha}\rangle=\frac{V_iC_\alpha}{W}$ and $\langle 
w_{i\alpha}^2\rangle=\left(\frac{V_iC_\alpha}{Wp_{i\alpha}}\right)^2p_{i\alpha}$, 
which leads to the expression
\begin{equation}
\frac{\langle 
S_i\rangle_{\text{ECAPM}}}{S_i}=\frac{\sum_\alpha\left(\frac{V_iC_\alpha}{Wp_{
i\alpha}}\right)^2p_{i\alpha}+\frac{V_iC_\alpha}{W}\left[C_\alpha-\frac{
V_iC_\alpha}{W}\right]}{\sum_\alpha w_{i\alpha}C_\alpha}.
\end{equation}

In the case of MECAPM instead we have $\langle 
w_{i\alpha}\rangle_{\text{MECAPM}}=\frac{V_iC_\alpha}{W}$ and $\langle 
w_{i\alpha}^2\rangle_{\text{MECAPM}}=\frac{V_iC_\alpha}{W}\left(1+2\frac{
V_iC_\alpha}{W}\right)$, which leads to the expression
\begin{equation}
\frac{\langle 
S_i\rangle_{\text{MECAPM}}}{S_i}=\frac{\sum_\alpha\frac{V_iC_\alpha}{W}\left[
1+C_\alpha+\frac{V_iC_\alpha}{W}\right]}{\sum_\alpha w_{i\alpha}C_\alpha}.
\end{equation}

Since, in both cases, the term $\frac{V_iC_\alpha^2}{W}$ dominates over the 
other ones and $\langle w_{i\alpha}\rangle_{\text{ECAPM}}=\langle 
w_{i\alpha}\rangle_{\text{MECAPM}}=\langle w_{i\alpha}\rangle_{\text{CAPM}}$, we have
\begin{eqnarray}\label{eqapm}
\frac{\langle S_i\rangle_{\text{ECAPM}}}{S_i}&\simeq&\frac{\langle 
S_i\rangle_{\text{MECAPM}}}{S_i}\simeq\frac{\sum_\alpha\frac{V_iC_\alpha^2}{W}}{
\sum_\alpha w_{i\alpha}C_\alpha}=\nonumber\\
&=&\frac{\sum_\alpha \langle 
w_{i\alpha}\rangle_{\text{CAPM}} C_\alpha}{\sum_\alpha w_{i\alpha}C_\alpha}.
\end{eqnarray}

Eq.(\ref{eqapm}) implies that, among the reconstruction methods satisfying the strength constraints, 
those which are expected to better reproduce the systemicness index are the ones better reproducing the observed 
weights. As discussed in the main text, the prescription $\langle 
w_{i\alpha}\rangle_{\text{ECAPM}}=\langle 
w_{i\alpha}\rangle_{\text{MECAPM}}=\frac{V_iC_\alpha}{W}$ is particularly 
successful in reproducing the largest weights of the SHS network (beside ensuring 
that $\langle C_\alpha\rangle=C_\alpha$).

\medskip

The statistical fluctuations affecting the $S_i$ index are
\begin{widetext}
\begin{equation}
\sigma^{2,\:\text{MECAPM}}_{S_i}=\sum_\alpha\langle 
w_{i\alpha}\rangle\Bigg[(1+2\langle w_{i\alpha}\rangle)\sum_j\langle 
w_{j\alpha}\rangle^2+C_\alpha(1+C_\alpha)+C_\alpha\langle 
w_{i\alpha}\rangle(2+C_{\alpha})\Bigg]
\end{equation}
\begin{equation}
\sigma^{2,\:\text{ECAPM}}_{S_i}=\sum_\alpha\frac{\langle 
w_{i\alpha}\rangle^2}{p_{i\alpha}}\Bigg[\sum_j\frac{\langle 
w_{j\alpha}\rangle^2}{p_{j\alpha}}+C_\alpha^2(1-p_{i\alpha})
-\sum_j\langle w_{j\alpha}\rangle^2\Bigg].
\end{equation}
\end{widetext}
Interestingly enough, plotting the vector of ratios 
\begin{equation}
r_{S_i}=\frac{\sigma_{S_i}^\text{ECAPM}}{\sigma_{S_i}^{\text{MECAPM}}}
\end{equation}
versus the nodes strengths $V_i$ reveals a functional dependency 
$r_{S_i}\propto V_i^{-1/2}$. Again, this confirms that ECAPM 
outperforms MECAPM in providing an estimate of the systemicness index 
$S_i$ for the larger institutions, which become less sensitive to statistical fluctuations.

\end{document}